\documentclass{aa}  

\usepackage{graphicx}
\usepackage{txfonts}
\usepackage[colorlinks]{hyperref}

\def\aap{A\&A}
\def\apj{ApJ}
\def\apjs{ApJS}

\def \hi {\ion{H}{i}}

\def\kms{km\,s$^{-1}$}

\begin{document}

\title{Correlations between turbulent velocity and density fields in the
  local interstellar medium }

   \subtitle{}

 \author{P.\ M.\ W.\ Kalberla \inst{1},  J.\ Kerp \inst{1} \and U.\ Haud \inst{2} }   

\institute{Argelander-Institut f\"ur Astronomie,
           Auf dem H\"ugel 71, 53121 Bonn, Germany \\
           \email{pkalberla@astro.uni-bonn.de}
           \and
           Tartu Observatory, University of Tartu,
           61602 T\~oravere, Tartumaa, Estonia }

\authorrunning{P.\,M.\,W. Kalberla, J. Kerp \& U.\ Haud }       

   \titlerunning{ISM turbulence in velocity and density}

%   \date{Received 17 September 2021 / Accepted 26 January 2022   }

  \abstract {Kalberla et al. used HI4PI data to analyze velocity and
    density fluctuations in the interstellar medium (ISM). They applied
    the \citet{Yuen2021} velocity decomposition algorithm (VDA) for
    separating such fluctuations in the position-position-velocity (PPV)
    space.  In the first version of this manuscript they came to the
    conclusion that velocity and density fields are statistically
    correlated. \citet{Yuen2021} tried to reproduce these results and
    pointed to a likely mistake in the VDA expression that was used. We
    confirm that there was such a software problem. The statement that
    VDA derived density and velocity fields from HI4PI are
    anti-correlated needs to be withdrawn. Correct is that these density
    and velocity fields are uncorrelated. In turn major parts of
    the conclusions in the first version of this manuscript, based
    on an erroneous correlation, are invalid. The submission to \aap\ was
    withdrawn on 24 February 2022.

  }

% -------------------------------

% -------------------------------
  \maketitle
%
%________________________________________________________________
 
\section{Introduction}
\label{Intro}

In the first version of the arXiv submission of this manuscript
\citep{Kalberla2022} we came to the conclusion that observed turbulent
HI4PI velocity and density fields, derived by using the \citet{Yuen2021}
velocity decomposition algorithm (VDA), are statistically
anti-correlated. This statement is incorrect.

\begin{figure*}[thp] %%  1
   \centering
   \includegraphics[width=6cm]{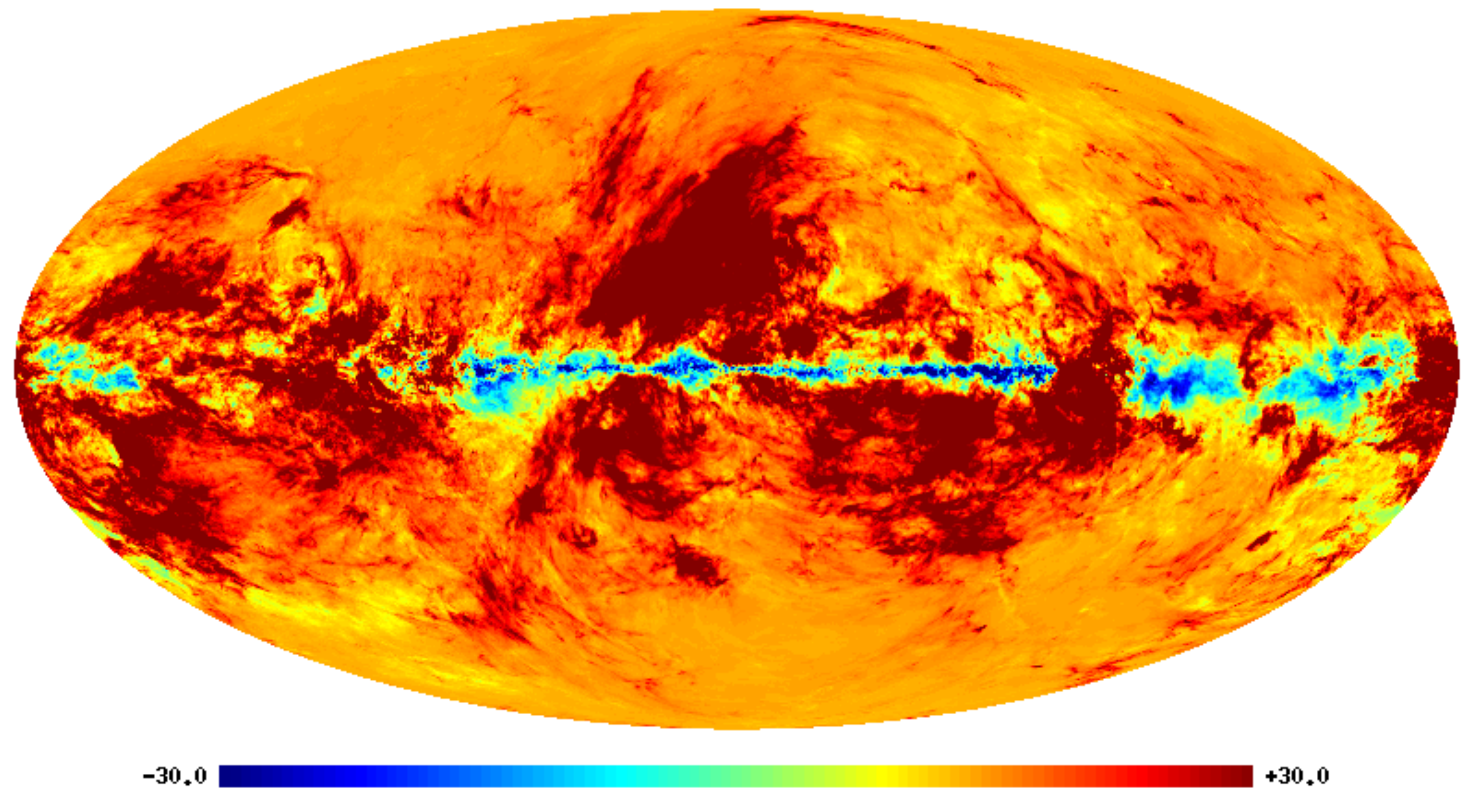}
   \includegraphics[width=6cm]{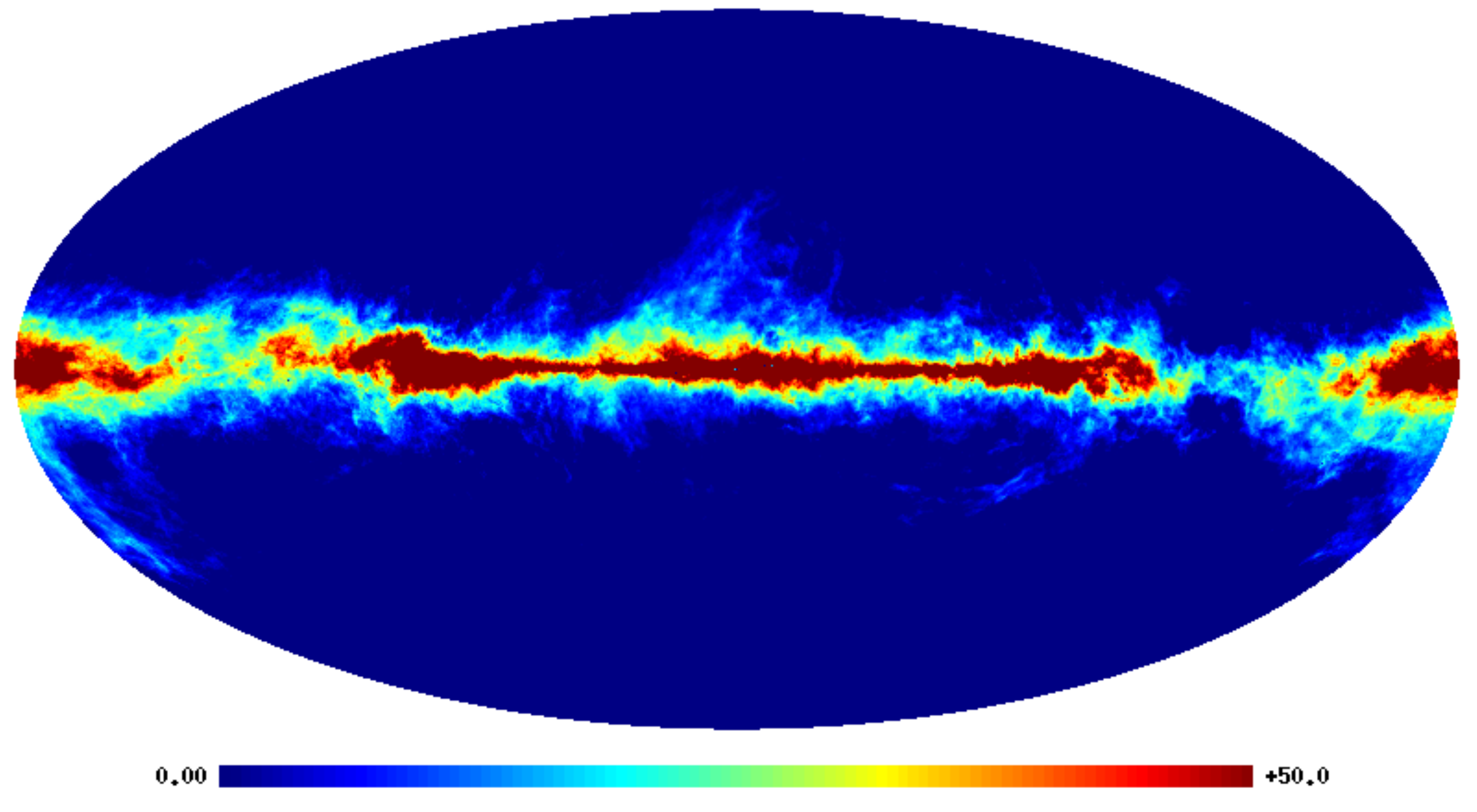}
   \includegraphics[width=6cm]{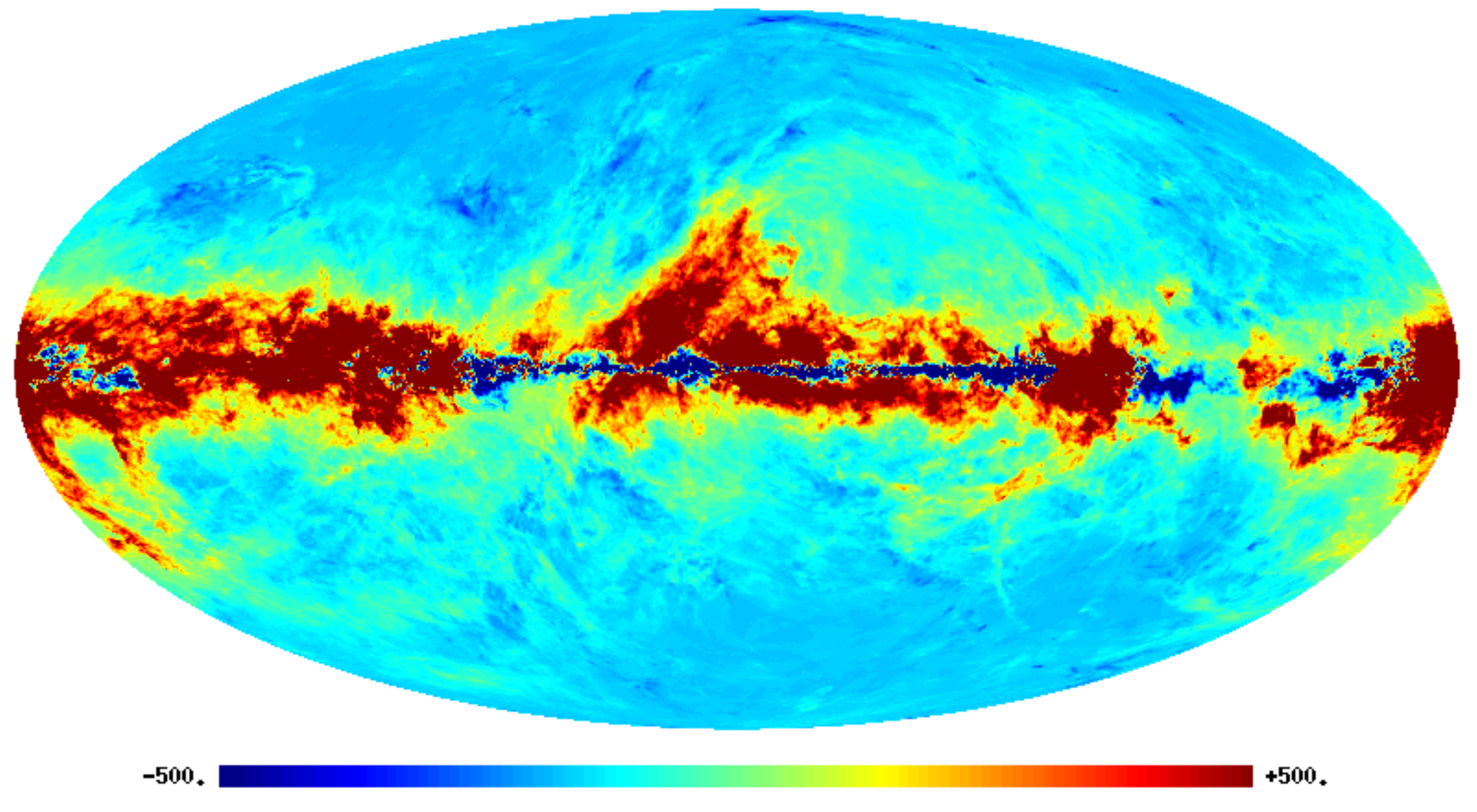}
   \caption{All-sky Mollweide displays for a single channel at
     $v_{\mathrm{LSR}} = 1 $ \kms. Left: derived velocity field $p_v$, middle:
     derived density field $p_d$, right: orthogonality relation
     $p_d p_v $ for $p_d$ and $p_v$. These plots have been generated for
     comparison with Fig. 1 of \citet{Yuen2022}. }
   \label{Fig_all_sky}
\end{figure*}

After our first version of manuscript became available Yuen et al. have
submitted a Letter to A\&A that got also available as
\citet{Yuen2022}. These authors contradicted our previous conclusions
and suggested with their Eq. (3) a possible systematical error that
invalidates our analysis. A check of the software has shown that there
was actually a mistake of this kind, leading to erroneous results and in
turn to invalid conclusions.

\begin{figure*}[thp] %%  1
   \centering
   \includegraphics[width=9cm]{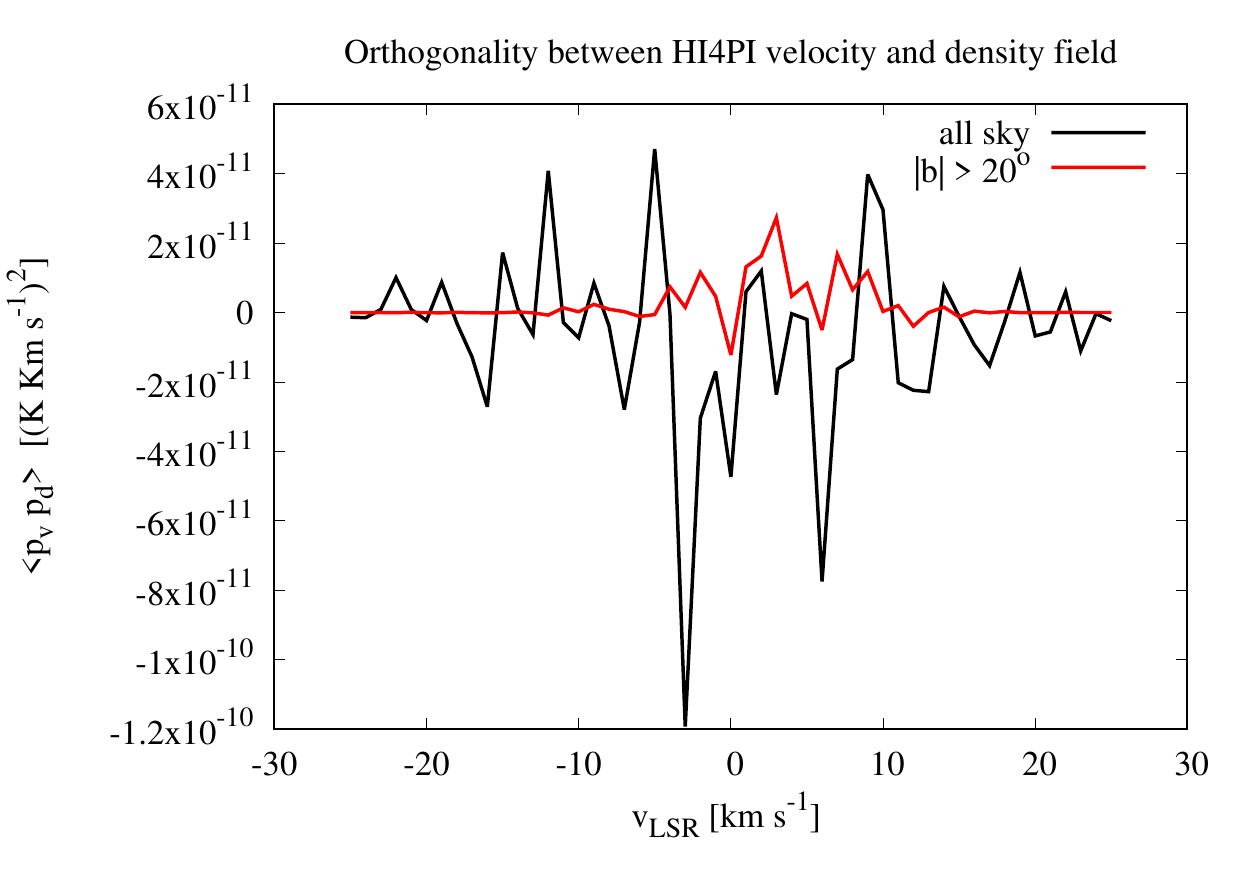}
   \includegraphics[width=9cm]{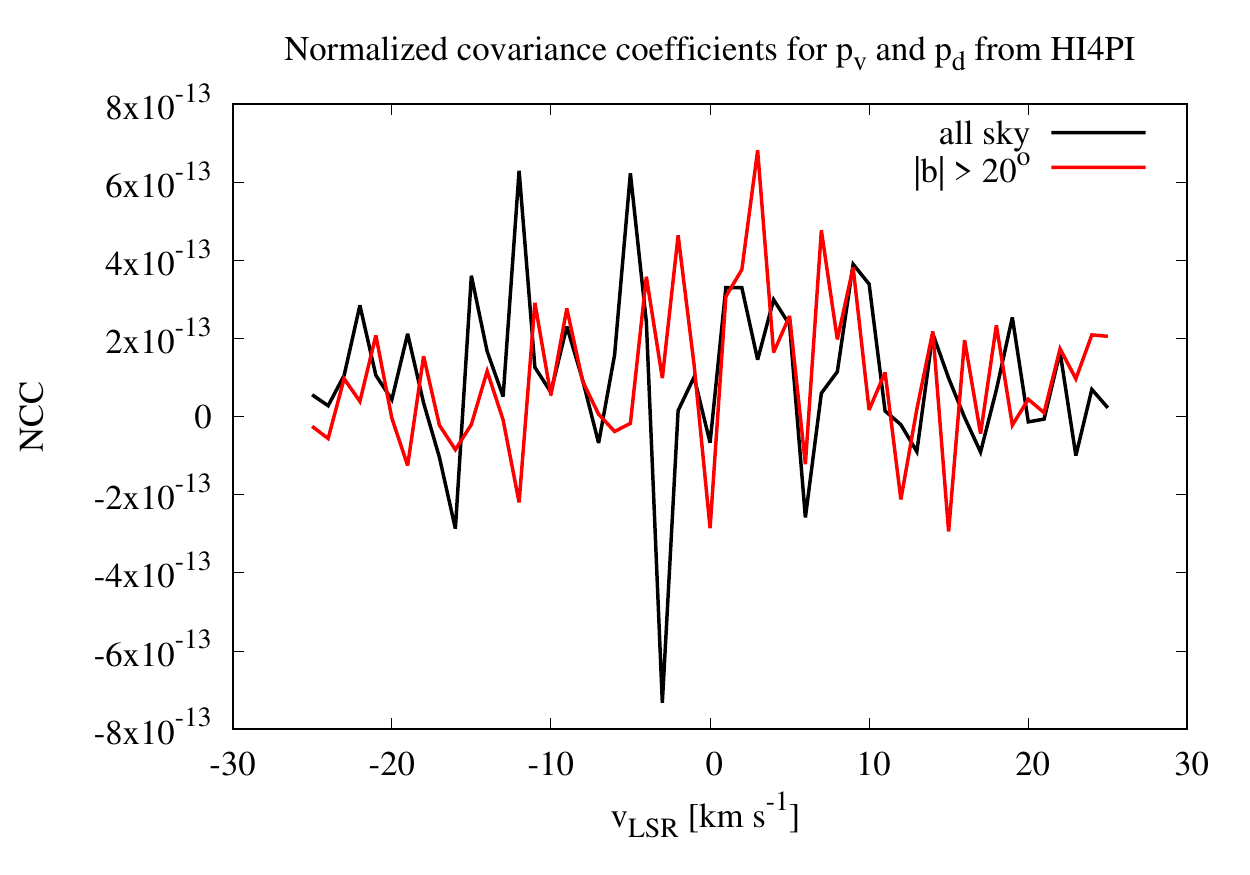}
   \caption{Left: Orthogonality relation $\langle p_v p_d \rangle $.
     Right: Normalized covariance coefficients derived from HI4PI data
     for the multiphase medium. The red line shows the high galactic
     latitude sky while the black one all-sky data. Both plots are for
     comparison with Fig. of \citet{Yuen2022}. }
   \label{NCC}
\end{figure*} 

\section{Updated data analysis}
\label{VDA}

After correction of this software problem we obtain significant
different results for the VDA derived channel maps. As in the first
version of this manuscript we consider a single velocity channel at
$v_{\mathrm{LSR}} = 1 $ \kms\ and total intensities integrated over the
velocity range $-25 < v_{\mathrm{LSR}} < 25 $ \kms, both derived from
Gaussian components. The derived VDA velocity field $p_v$, the density
contribution $p_d$, and the product $p_d p_v $ are shown in
Fig. \ref{Fig_all_sky} with similar scaling as used by \citet{Yuen2022}
in their Fig 1.

Based on VDA velocity and density channelmaps, generated with the
updated software, we calculated the orthogonality relation $\langle p_v
p_d \rangle $ and normalized covariance coefficients $NCC(p_v,p_d)$
according to Eq. (21) of \citet{Yuen2021} for multiphase HI4PI data. The
results are shown in Fig. \ref{NCC} and should be compared with Fig. 2
of \citet{Yuen2022}.  It is obvious that any correlations between
velocity and density fields are missing. The previous finding that there
is a strong anti-correlation is based on a software failure. All
conclusions in the first version of our submission that are based on an
anti-correlation are invalid.  From HI4PI data we find that the observed
multiphase \hi, also all phases individually with any possible field
selections in Galactic coordinates satisfy $\langle p_v p_d \rangle = 0
$ and $NCC(p_v,p_d) = 0$, independent of the velocity interval selected
for the analysis.

%=========================================================================
%=========================================================================
%=========================================================================
%=========================================================================
%=========================================================================
%=========================================================================
\begin{acknowledgements}
The correction and withdrawal of our first version would not have been
possible without the advanced publication of the Yuen et al. Letter in
the arXiv research-sharing platform. EBHIS is based on observations with
the 100-m telescope of the MPIfR (Max-Planck-Institut f\"ur
Radioastronomie) at Effelsberg. The Parkes Radio Telescope is part of
the Australia Telescope which is funded by the Commonwealth of Australia
for operation as a National Facility managed by CSIRO. Some of the
results in this paper have been derived using the HEALPix package.
\end{acknowledgements}

\end{document}